\documentclass[sigconf,nonacm]{acmart2}
\usepackage{bbm}
\AtBeginDocument{%
  }

\setcopyright{acmcopyright}
\copyrightyear{2023}
\acmYear{2023}
\acmDOI{XXXXXXX.XXXXXXX}

\usepackage[subrefformat=parens]{subcaption}

\begin{document}

\title{Adversarial Deep Hedging: Learning to Hedge\\ without Price Process Modeling}

\author{Masanori HIRANO}
\orcid{0000-0001-5883-8250}
\affiliation{%
  \institution{The University of Tokyo}
  \city{Tokyo}
  \country{Japan}
}
\email{research@mhirano.jp}

\author{Kentaro MINAMI}
\affiliation{%
  \institution{Preferred Networks, Inc.}
  \city{Tokyo}
  \country{Japan}
}
\email{minami@preferred.jp}

\author{Kentaro IMAJO}
\affiliation{%
  \institution{Preferred Networks, Inc.}
  \city{Tokyo}
  \country{Japan}
}
\email{imos@preferred.jp}

\renewcommand{\shortauthors}{Hirano et al.}

\begin{abstract}
Deep hedging is a deep-learning-based framework for derivative hedging in incomplete markets.
The advantage of deep hedging lies in its ability to handle various realistic market conditions, such as market frictions, which are challenging to address within the traditional mathematical finance framework.
Since deep hedging relies on market simulation, the underlying asset price process model is crucial.
However, existing literature on deep hedging often relies on traditional mathematical finance models, e.g., Brownian motion and stochastic volatility models, and discovering effective underlying asset models for deep hedging learning has been a challenge.
In this study, we propose a new framework called adversarial deep hedging, inspired by adversarial learning.
In this framework, a hedger and a generator, which respectively model the underlying asset process and the underlying asset process, are trained in an adversarial manner. 
The proposed method enables to learn a robust hedger without explicitly modeling the underlying asset process.
Through numerical experiments, we demonstrate that our proposed method achieves competitive performance to models that assume explicit underlying asset processes across various real market data.
\end{abstract}





\maketitle

\section{Introduction}
Derivative hedging is an important issue in the field of finance.
Practical derivative hedging is strongly connected to the theory of mathematical finance, which was pioneered by the work of Black, Scholes and Merton in the 1970s \cite{black1973,merton1973}.
In mathematical finance, the financial market is typically modeled as a complete market, i.e., an ideal market with continuous tradability and no friction.
Under this assumption, the optimal hedging strategy is a portfolio that replicates the derivative, which also leads to the theoretical price of the derivative from the no-arbitrage argument.
However, the assumption of the complete market does not hold in reality, and perfect hedging is impossible.
Therefore, practical hedging strategies are manually adjusted by human traders using various empirical rules.
Greeks are the most typical tools for managing risk sensitivities.
By managing greeks, traders aim for robust hedging against market changes (see e.g., \cite{taleb1997dynamic}).

Deep hedging \cite{deep-hedging} was developed to address the problem of hedging in incomplete markets.
Deep hedging is a fully computational and data-driven approach based on market simulation and deep hedging techniques that directly optimizes the profit and loss (PL) of the portfolio of derivatives.
The advantage of deep hedging is that it can calculate the optimal hedging strategy in terms of a given utility function, while considering various characteristics of real markets, such as market friction, which are difficult to handle within the traditional framework of mathematical finance theory.

The choice of the market simulator, which simulates the price process of the underlying asset, is crucial in deep hedging training.
As deep hedging aims to learn the optimal hedging strategy on a given market simulator, the market simulator should cover all the possible scenarios in real markets.
Currently, in the literature on deep hedging, there are mainly two approaches to market simulator design:

\noindent\underline{(i) Models from mathematical finance}:
While deep hedging can be trained on any market simulator, many existing studies tend to conduct experiments based on underlying asset models developed in mathematical finance.
These models include geometric Brownian motion \cite{black1973,merton1973}, Heston model \cite{Heston1993}, rough volatility model \cite{Bayer2015}, and jump-diffusion model \cite{Cox1976,Merton1976}, among others.
This approach of utilizing established mathematical finance models is valid as it is both simple to implement and provides a fair comparison with traditional hedging techniques.
However, the validity of these models is hindered by their limited ability to simulate market scenarios that could happen in reality. 
One reason is that the traditional models aim to represent market stylized facts (e.g., volatility surfaces \cite{gatheral2012volatility}) using just a handful of parameters, while they are \textit{not} necessarily targeting market prediction or path simulations.
Additionally, as financial markets are subject to non-stationarity and regime shifts, it is not always possible to make good predictions with a single model. 
In situations where the assumed underlying asset price process deviates significantly from the actual market conditions, there is a risk of deteriorating hedge performance.

\noindent\underline{(ii) Machine-learning-based models}:
An alternative approach is to utilize machine learning methods to learn data-driven market simulators.
In recent years, the accelerated development of machine-learning-based generative models has facilitated this approach.
Some examples include Generative Adversarial Networks (GAN) training using the actual data in financial markets \cite{Li2020,Naritomi2020,hirano2022policy,wiese2020quant}, but which requires plenty of computational resources just for generating price series.
One of the most pressing challenges for machine-learning-based approaches is that the historical data from financial markets is comparably limited to the amount of data required by contemporary machine learning algorithms.
In response to this challenge, various techniques based on market stylized facts or data augmentation have been proposed to mitigate the effects of limited data.
For example, Ziyin {\it et al.} \cite{Ziyin2022-uq} proposed a new noise injection method for data augmentation in financial market data.
Nonetheless, it remains uncertain whether adjustments based on stylized facts are effective for the downstream task of deep hedging.

In this paper, we propose a novel alternative approach by adopting an adversarial environment to learn the hedging strategy.
Our proposed method consists of two adversarial neural networks: a hedger and a generator.
The hedger learns to maximize the utility function to optimize the hedging strategy, similar to the existing deep hedging framework.
On the other hand, the generator, which is newly introduced in our framework, is a generative neural network that simulates the underlying asset.
The generator learns to simulate the market to make it more difficult for the hedger to maximize the utility function.
The proposed method can be considered as a framework similar to adversarial learning.
It is expected to promote robust learning against environmental changes by a generator providing an environment that is unfavorable to the hedger.
This is a desirable property, considering that real markets are non-stationary and difficult to predict, and regime shifts exist.
Moreover, the modeling aspect of our approach has a distinct advantage in that it can learn the hedging strategy without presuming an explicit underlying asset model.
Although our proposed approach is among the earliest attempts at adversarial hedging, we demonstrate through numerical experiments in Section \ref{sec:experiment} that the adversarially learned hedger performs well in real markets.

\if0
The remainder of this paper is organized as follows.
In Section \ref{sec:related}, we review related literature on deep hedging.
In Section \ref{sec:task}, we formulate the problem setting of hedging options.
In Section \ref{sec:method}, we introduce our proposed framework, adversarial deep hedging, in an abstract level.
In Section \ref{sec:num-analysis}, we discuss the convergence property of adversarial deep hedging using numerical analysis on a toy example.
In Section \ref{sec:experiment}, we present the experimental setting on real-world market data, followed by discussion on the results in Section \ref{sec:result} and Section \ref{sec:discussion}.
Section \ref{sec:conclusion} concludes.
\fi

\section{Related Work}\label{sec:related}
\if0
In terms of option hedging and prices, mathematical finance approaches have been frequently used.
The celebrated Black--Scholes model \cite{black1973,merton1973} is certainly the most famous model in the option pricing literature.
To date, the Black--Scholes model has been widely used in practice, whose applications include deriving option pricing formulae (e.g., the Black--Scholes formula) and analyzing structures of option prices (e.g., greeks).

Advances in neural networks have opened up a new avenue of research on nonparametric approaches to option pricing and hedging.
Hutchinson {\it et al.} \cite{Hutchinson1994} and Garcia {\it et al.} \cite{Garcia2000} showed that multi-layer perceptron (MLP) could replace the Black--Scholes model.
Malliaris {\it et al.} \cite{Malliaris1996} showed that neural networks could predict implied volatility by using the index option of the S\&P100.
For a comprehensive review, see Ruf and Wang \cite{Ruf2020} and references therein.
\fi

In their seminal work \cite{deep-hedging,Buehler2019b}, Buehler {\it et al.} proposed a method called \emph{Deep Hedging}.
Deep hedging utilizes neural networks to model hedging strategies and construct optimal strategies by directly maximizing utility functions derived from the final Profit \& Loss (PL).
Deep hedging has been recognized as a breakthrough in the derivative industry, owing to two prominent features:
First, deep hedging can handle various types of frictions by a fully computational approach.
Second, deep hedging bypasses the need for theoretical pricing models, which are prerequisites for other standard hedging methods, such as those based on Greeks.
Since the introduction of deep hedging, many efforts have been devoted to expand its versatility.
Imaki {\it et al.} \cite{Imaki2023} proposed a deep hedging model that incorporates no-transaction band \cite{Davis1993} into the neural network architecture, resulting in fast convergence of learning.
Murray {\it et al.} \cite{Murray2022-nz, Buehler2022-nr} proposed an actor-critic-based reinforcement learning for deep hedging and achieved almost equivalent performance to the original deep hedging.
Hirano {\it et al.} \cite{Hirano2023-iiaiaai-ndh} extended the deep hedging framework into a nested structure, thereby facilitating the use of options as hedging instruments. The authors also proposed efficient simulation designs and a learning algorithm to circumvent the computational issues.

Several recent studies focus on the data used for learning of deep hedging.
Mikkila {\it et al.} \cite{Mikkila2023-gk} argued that deep hedging often depends on a specific choice of price process simulator, which could result in suboptimal performance. They suggested that improved performance could be achieved by utilizing empirical data for training. However, the amount of available historical market data is often limited. 
Horvath {\it et al.} \cite{Horvath2021} proposed a deep hedging model based on rough volatility models such as the rBergomi model \cite{Bayer2015}, which can model (non-Markov) jumps in price processes.

Other studies have also focused on generating financial time series using deep-learning-based models.
For example, several researchers have proposed variants of Generative Adversarial Networks (GANs) for generating financial time series \cite{Li2020,Naritomi2020,hirano2022policy,wiese2020quant}.
Hayashi {\it et al.} \cite{Hayashi2022-fr} proposed a generative diffusion model tailored for financial markets.
While these were originally intended for purposes other than deep hedging, they can be adapted for use within the deep hedging framework.
However, to the best of our knowledge, there has not been ample exploration of deep generative models explicitly designed with the direct goal of enhancing deep hedging performance. This specific research gap is what we aim to address in this paper.

\if0
Among numerous simulators for price processes, it is extremely challenging to determine which one is appropriate or whether a simulator should not be used. Therefore, this study focuses on deep hedging free from the assumption of underlying asset price processes.
\fi

\section{Setting: Discrete Time Market with Frictions}\label{sec:task}
Throughout this paper, we consider the problem of hedging an option with its underlying asset.
We consider a discrete-time financial market with a finite time horizon $T$ and trading dates $0=t_0 < t_1 < \cdots < t_n = T$.
At $t=t_0$, the hedger (the issuer of the option) sells one unit of the option.
The hedger aims to hedge the risk arising from the future payoff of the option by trading the underlying asset $S$.
$S$ is assumed to be always tradable, and its price at time $t$ is denoted as $S_t$.
At each trading date $t_i$, the hedger sets a new position $\delta_{t_i}$ in $S$.
For all transactions of $S$, a proportional transaction cost is charged at a rate of $c$ per unit price.

Option $Z$ is a contract in which the payoff at time $t = T$ depends on the price of the underlying asset $S$.
Examples include an European option $Z(S) = \max(S_T-K, 0)$ and a Lookback option $Z(S) = \max(\max(S)- K, 0)$, where $K$ denotes the strike value of the options.
These options will also be used in the experiment in Section \ref{sec:experiment}.

The Profit \& Loss (PL) of the hedger (the issuer) is formulated as follows:
\begin{eqnarray}
	\mathrm{PL}_T (Z, S, \delta) &:=& -Z(S) + (\delta\cdot S)_T - C_T (\delta)\label{eq:pl}\\
	(\delta\cdot S)_T &:=& \sum_{i=0}^{n-1} \delta_{t_i} (S_{t_{i+1}} - S_{t_i}) \\
	C_T (\delta) &:=& \sum_{i=0}^{n} cS_{t_i}|\delta_{t_i} - \delta_{t_{i-1}}|
\end{eqnarray}
where $\delta_{-1} = \delta_{t_n} = 0$.
Here, $(\delta\cdot S)_T$ is the total return resulting from the trading of the underlying asset, and $C_T (\delta)$ is the total transaction costs.

The objective function of the hedger is defined by the PL and the utility function $u$.
Specifically, the loss function to be minimized is defined as follows:
\begin{eqnarray}
	l(\delta) = - u \left(\mathrm{PL}_T(Z, S, \delta)\right).
\end{eqnarray}
For an example of the utility function $u$, Entropic Risk Measure (ERM) is defined as
\begin{eqnarray}
	u(x) = -\frac{1}{\lambda}\log\mathbbm{E}\left[\exp{(-\lambda x)}\right]\label{eq:erm}
\end{eqnarray}
where $\lambda$ is the risk preference coefficient.
Expected Shortfall (Conditional Value at Risk; CVaR) is defined as
\begin{eqnarray}
	u(x) = \frac{1}{1-\alpha} \int_{-\infty}^{\mathrm{VaR}(\alpha)} x\cdot p(x)dx\label{eq:cvar}
\end{eqnarray}
where $\alpha$ is the confidence level and satisfies $0\leq\alpha\leq 1$.

\section{Method: Adversarial Deep Hedging}\label{sec:method}

Our goal is to learn deep-learning-based hedging strategies that perform well in real-world markets.
As we mentioned in the introduction, an efficient design of the market simulator is crucial to improve the performance of deep hedging.
While the existing studies of deep hedging often employ underlying asset price models from the mathematical finance literature,
we here study more flexible deep-learning-based models to pursue better training of hedging strategies.

The central idea of our approach is to employ the concept of \textit{adversarial learning}.
In the deep learning literature, it is well known that there are ``difficult'' input data called adversarial examples
that significantly degrade the performance on the machine learning tasks such as image classification \cite{Goodfellow2014-rd,Madry2017-am}.
Adversarial examples can be obtained by adding noises to existing images in training dataset.
The noises are optimized toward the worst-case directions that deteriorates the classicication performance most
(hence the name of ``adversarial'' examples).
Interestingly, it is shown that adding adversarial noises during training improves the overall classification performance \cite{Goodfellow2014-rd}.
This idea would be useful for improving the market model for deep hedging training.
As real market data can contain many types of price movements that are difficult to be modeled (e.g., jumps and regime shifts),
we expect that adversarial learning can make the deep hedging model robust against unseen environmental changes
that are never generated from a simple market simulator.

To formulate the idea of adversarial learning of deep hedging, we propose a framework based on the mim-max game paradigm,
akin to Generative Adversarial Networks (GANs) \cite{Goodfellow2014}.
This approach offers several potential advantages.
Firstly, it is anticipated that the hedger is robustly trained against perturbations in the market environment.
Secondly, we can circumvent the need for explicit modeling of the asset model.

\begin{figure}[tbp]
	\centering
	\includegraphics[width=1.0\linewidth]{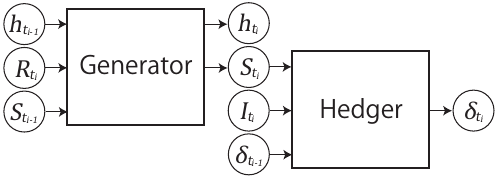}
	\caption{Inputs/outputs of adversarial deep hedging. $h_{t_i}$ is a hidden state of the generator, $R_{t_i}$ is a random variable for generation, $S_{t_i}$ is underlying asset price, $I_{t_i}$ is available information for the hedger, and $\delta_{t_i}$ is the hedge position.}
	\label{fig:adversarial-flow}
\end{figure}

Figure \ref{fig:adversarial-flow} shows the outline of the adversarial deep hedging.
Similar to the original deep hedging, the hedger receives the underlying asset price $S_t$ and relevant information as inputs,
and determines the optimal hedging position.
We introduce a new component, the generator, that handles market simulation.

In what follows, we will elaborate on each component of our proposed framework.

\subsection{Hedger}
Hedger $H$ is a neural network that generates a new hedge position $\delta_{t_i}$ using the relevant information at time $t_i$:
\begin{eqnarray}
    \delta_{t_i} = H( S_{t_i}, I_{t_i}, \delta_{t_{i-1}}),
    \label{eq:hedger}
\end{eqnarray}
where $S_{t_i}$ is underlying asset price, $I_{t_i}$ is other relevant information at time $t_i$, and $\delta_{t_{i-1}}$ is the previous hedge position.
$I_{t_i}$ can contain information related to theoretical optimal hedging, such as Black--Scholes delta, as well as any information of market conditions.
Since the theoretical optimal hedging depends on the previous position in the presence of transaction costs (see e.g., \cite{Imaki2023}), it is natural for the hedger to take $\delta_{t_{i-1}}$ as an input.

\subsection{Generator}
Generator $G$ is a neural network generating an underlying asset price sequence.
As the generator should generate causal time series,
it is sensible to employ Recurrent Neural Network (RNN)-type architecture, as described below.
The generator has hidden state vector $h_{t}$.
At each time step $t_i$, the generator receives the previous hidden state $h_{t_{i-1}}$, asset price $S_{t_{i-1}}$, and a random variable $R_{t_i}$, and outputs a new asset price $S_{t_i}$ and hidden state $h_{t_i}$:
\begin{eqnarray}
    (S_{t_i}, h_{t_i}) = G( h_{t_{i-1}}, S_{t_{i-1}}, R_{t_i}).
    \label{eq:generator}
\end{eqnarray}
Here, similar to the generator in a standard GAN, $R_{t_i}$ plays a role in injecting exogeous randomness for generation.

\subsection{Adversarial Learning}

In the spirit of GAN, the adversarial deep hedging is formulated as a min-max game.
More formally, the hedger and the generator are trained through the following optimization problem:
\footnote{
Here, to enhance the intuitive understanding of functional dependence, we omitted the time index $t$ and used simplified notations. The generator in Eq.~\eqref{eq:generator} and the hedger in Eq.~\eqref{eq:hedger} are abbreviated as $G(R)$ and $H(G(R), I)$, respectively.
}
\begin{eqnarray}
	\min_G \max_H u\left(\mathrm{PL}_T(Z, G(R), H(G(R), I))\right).\label{eq:min-max}
\end{eqnarray}
The generator and the hedger are trained in an alternating and adversarial manner.
Specifically, the hedger strives to maximize the utility function of the terminal wealth,
while the generator aims to adversarially learn the distribution of unfavorable paths for the hedger.

\section{Numerical Stability Analysis}\label{sec:num-analysis}
In the previous section, we have introduced the concept of adversarial deep hedging at an abstract level.
However, it is not clear whether the min-max problem in Eq.~\eqref{eq:min-max} is well-posed.
In other words, it is not evident whether there exists an optimal solution or a local equilibrium, nor is the stability of learning guaranteed in practical settings.
In fact, the well-posedness depends on the specific problem formulation, including the choice of the generator class, the hedger class, the option payoff, and the utility function.
For example, when the expressive power of the generator class is too high, it may allow for the generation of unrealistic and extreme paths, which could destabilize hedger learning.
Therefore, it is presumed that the learning stability requires some form of regularity assumptions to ensure learning stability.
Moreover, it is known that min-max problems like GANs often suffer from learning instability, which requires careful design of the learning algorithm.

To gain an intuitive understanding of learning stability, we here examine a toy problem described as follows.
We consider a market model with only one underlying asset $S_t$ and two time instants $t \in \{ 0, 1 \}$
(i.e., $n = 1$ in Section \ref{sec:task}).
We assume that $S_0$ is a given constant.
Thus, the generator $G$ is responsible for generating a single price $S_1 \in \mathbb{R}$, and the hedger $H$ determines a scalar-valued position $\delta \in \mathbb{R}$.
The payoff function of the option is given as $Z(S_1) = \max\{ S_1 - K, 0\}$, where the strike value $K$ is assumed to be same as the initial price $S_0$.
Then, the PL at the terminal time is written as:
\begin{eqnarray}
    \mathrm{PL}_T (Z, S_1, \delta) = -Z(S_1) + \delta (S_1 - S_0) - c |\delta| S_0.\label{ref:anl-pl}
\end{eqnarray}

For simplicity, we assume that the distribution of $S_1$ is a Gaussian distribution
$
	S_1 - S_0 \sim \mathcal{N}(\mu, \sigma),
$
where the generator determines $\mu$ and/or $\sigma$.

For the utility function, we will use ERM(10.0) and CVaR(0.95).
Here, ERM(10.0) means the ERM-based utility function with $\lambda=10.0$ in Eq. (\ref{eq:erm}), and CVaR(0.95) means the CVaR-based utility function with $\alpha=0.95$ in Eq. (\ref{eq:cvar}).

\subsection{Case 1: Fixed Generator}
Firstly, we fixed the generator parameters $(\mu, \sigma)$ and checked the response of $\delta$ (the hedger) to the utility.
We set $\mu = 0.0, \sigma=0.2$, and the goal is to maximize the utility function with respect to $\delta$.

\begin{figure}[tbp]
	\centering
	\includegraphics[width=0.49\linewidth]{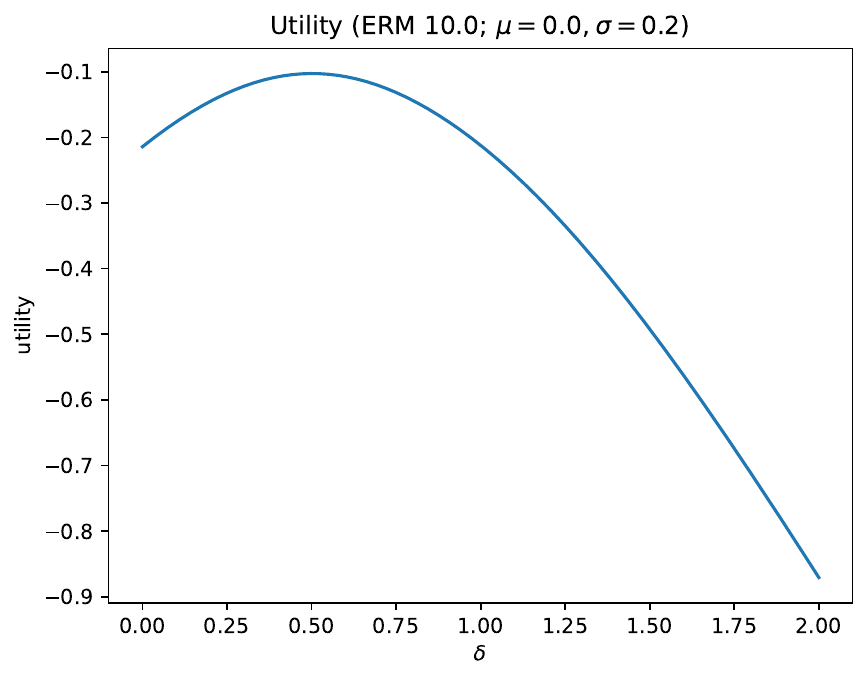}
    \includegraphics[width=0.49\linewidth]{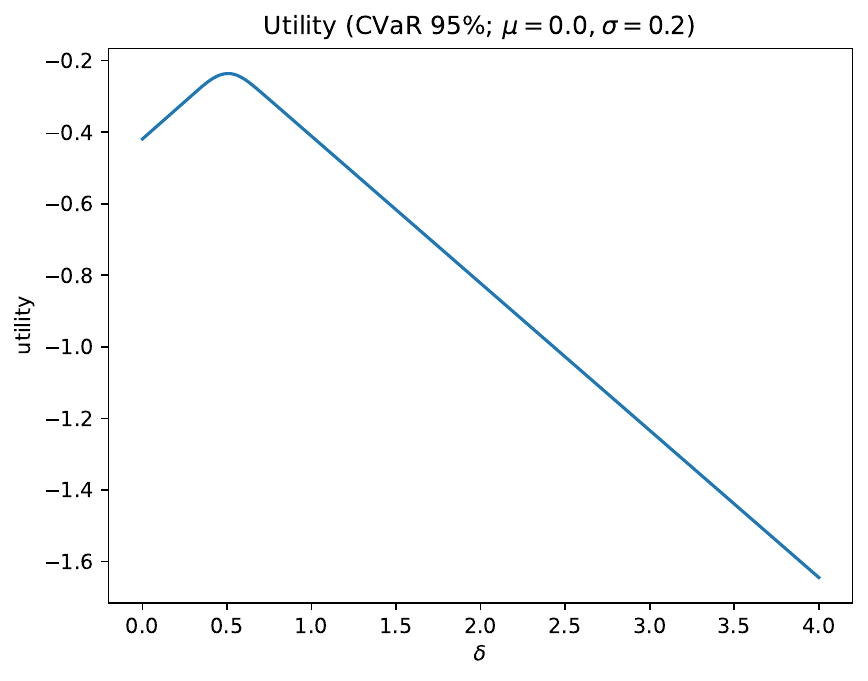}
	\caption{Results (left: ERM(10.0), right: CVaR(0.95)) of Utility vs. $\mathbf{\delta}$ under the Generator is fixed. Peaks exist at $\mathbf{\delta = 0.5011}$ for EMR(10.0) and $\mathbf{\delta = 0.5082}$ for CVaR(0.95).}
	\label{fig:fixed_gen}
\end{figure}

We plot the utility values in Figure \ref{fig:fixed_gen}.
It is evident that each utility function is concave in $\delta$, and the optimization problem has a unique optimal solution.
In each case, the utility function exhibits a peak around $\delta = 0.5$.
This means that the optimal solution is nearly identical to the Black--Scholes' delta hedging.
However, owing to the cost (i.e., the last term in Eq. (\ref{ref:anl-pl})), the utility peak is slightly different from $\delta=0.5$.

\subsection{Case 2: Adversarially Optimized Mean}\label{sec:anal2}
Secondly, we tested the case where the generator can adversarially optimize the mean parameter $\mu$.

\begin{figure}[tbp]
	\centering
	\includegraphics[width=0.4\linewidth]{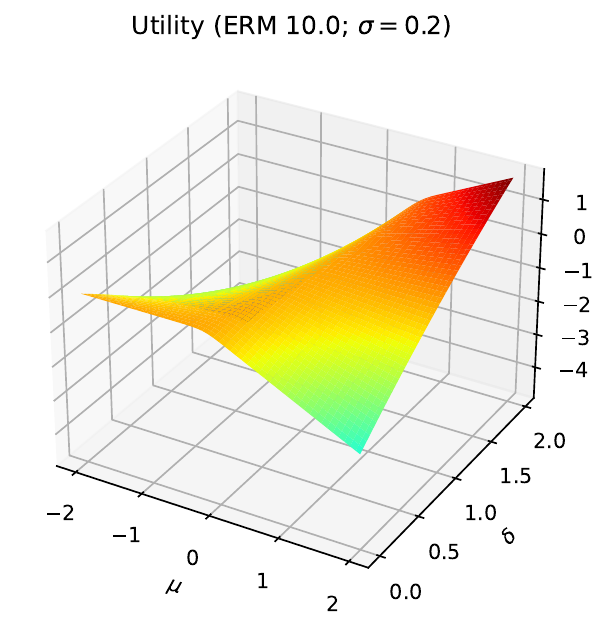}
    \includegraphics[width=0.59\linewidth]{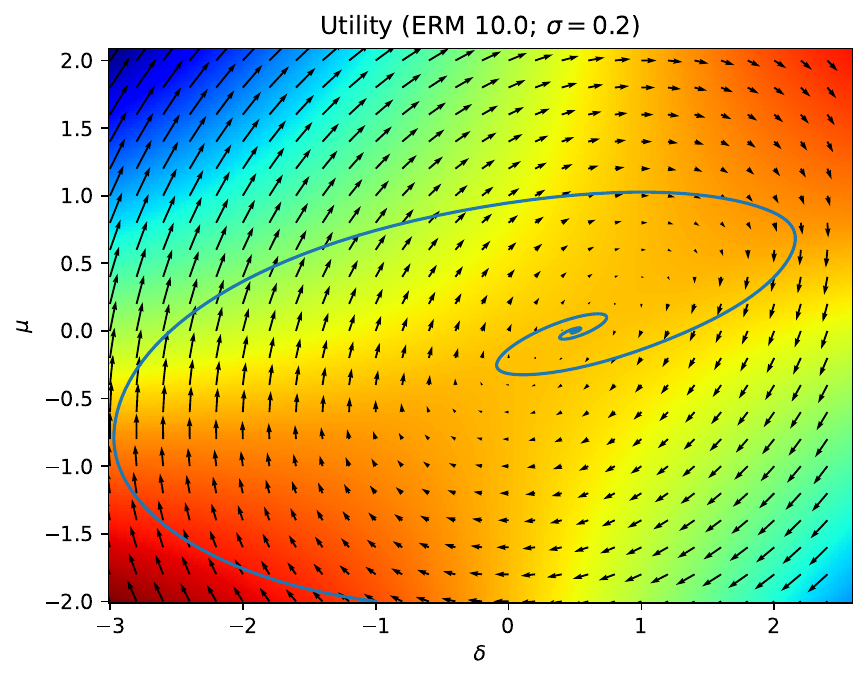}
	\caption{Results of utility among different $\delta$ and $\mu$ (ERM(10.0)). The left figure shows the utility surface and the right figure shows the utility heatmap and gradient arrow graph, and the blue line is the result of adversarial training simulation.}
	\label{fig:fixed_sigma_erm}
\end{figure}

\begin{figure}[tbp]
	\centering
	\includegraphics[width=0.4\linewidth]{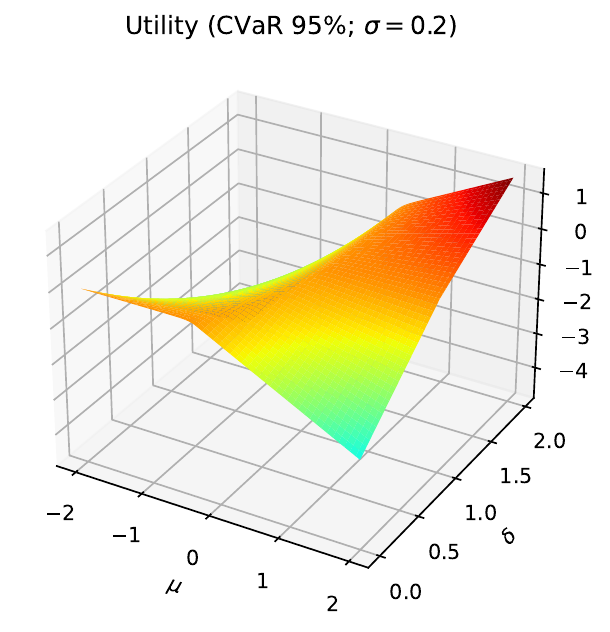}
    \includegraphics[width=0.59\linewidth]{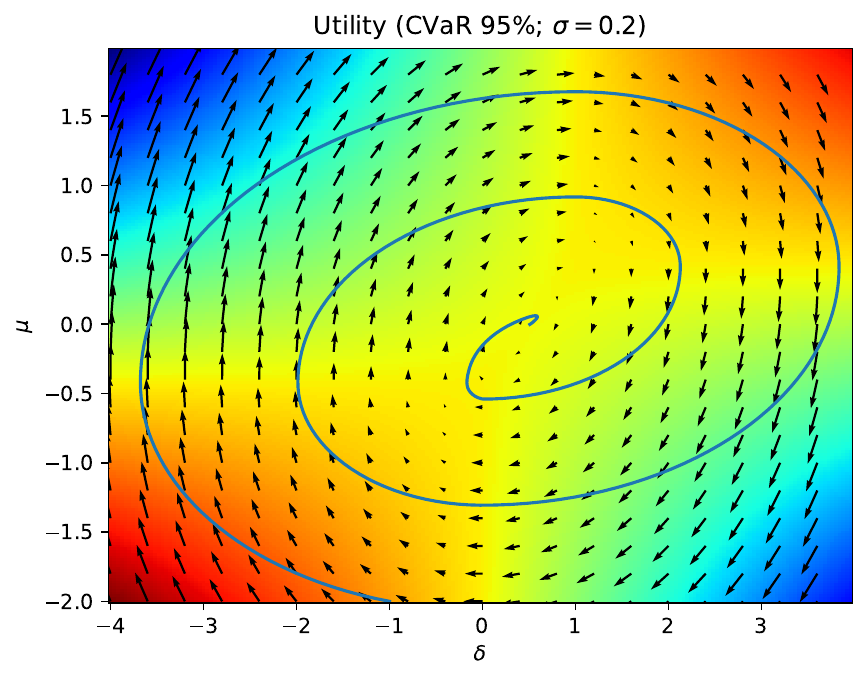}
	\caption{Results of utility among different $\delta$ and $\mu$ (CVaR(0.95)). The left figure shows the utility surface and the right figure shows the utility heatmap and gradient arrow graph, and the blue line is the result of adversarial training simulation.}
	\label{fig:fixed_sigma_cvar}
\end{figure}

Figure \ref{fig:fixed_sigma_erm} and \ref{fig:fixed_sigma_cvar} show the utility surfaces and their gradients plotted against $\delta$ and $\mu$, respectively for ERM and CVaR.
We also run adversarial learning simulations and plotted their trajectories in the right panels.
For the configuration of the learning algorithm (e.g., the learning rates and the two time-scale update rule), we used the same settings as the experiments in Section \ref{sec:experiment}.

According to Figures \ref{fig:fixed_sigma_erm} and \ref{fig:fixed_sigma_cvar}, the min-max game formulated in Eq. (\ref{eq:min-max}) can be convergent into the point around $\delta=0.5, \mu=0.0$.
This suggests that, when the generator optimizes only the mean $\mu$ in an adversarial manner, the dynamical system determined by the learning algorithm can converge to some equilibrium.
Furthermore, this result also suggests that the most unfavorable market environment for the hedger entails a zero expected return of the underlying asset, which is intuitively understandable as the no-arbitrage condition.

\subsection{Case 3: Adversarially Optimized Volatility}
\begin{figure}[tbp]
	\centering
	\includegraphics[width=0.49\linewidth]{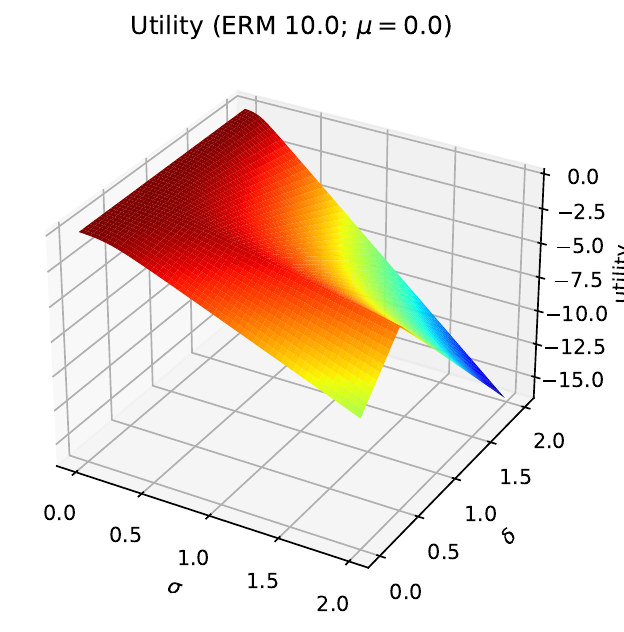}
    \includegraphics[width=0.49\linewidth]{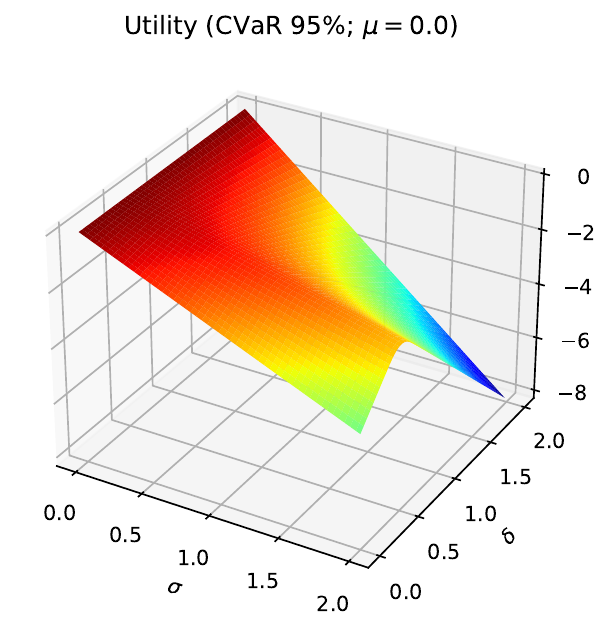}
	\caption{Utility surfaces among different $\delta$ and $\sigma$ (left: ERM(10.0), right: CVaR(0.95)).}
	\label{fig:fixed_mu}
\end{figure}

Thirdly, we fixed $\mu = 0$ and let the generator optimize the volatility parameter $\sigma$.
Figure \ref{fig:fixed_mu} shows utility surfaces plotted against $\delta$ and $\sigma$.

In either case, the utility value increases as $\sigma$ becomes larger, indicating that the Nash equilibrium of the min-max game does not exist.
As such, it is anticipated that the volatility can become arbitrarily large through the adversarial learning.
This is an intuitive result, since a more volatile market environment is more unfavorable for the hedger.
This example suggests that using a generator class with excessively high expressive power can lead to instability in adversarial learning.

Note that, for each fixed value of $\sigma$, $\delta$ always has a local minimum around $0.5$.
This suggests the hedger may converge to an optimal solution, even when the generator fails to converge.
One reason for this is that, in this particular example, the optimal hedging action remains unchanged regardless of the magnitude of the volatility.
In the experiment in Section \ref{sec:experiment}, we confirm that such ``scale invariance'' holds even in more realistic examples of European and lookback options.
In general settings, however, it is considered that the scale invariance does not hold, and it would be more reasonable to introduce some constraints on the volatility.
See a related discussion in Section \ref{sec:discussion-vol} as well.

\section{Experiment}\label{sec:experiment}
We conducted an experiment to demonstrate the effectiveness of our methods on real market data.
We used ten different experimental settings consisting of two options and five utility functions as follows: 
\begin{itemize}
	\item \textbf{Option $Z$}: We used (i) European Call Option and (ii) Lookback Call Option explained in Section \ref{sec:task}. For both types of options, the strike $K$ is set as $K (= S_{t_0}) = 1.0$.
	\item \textbf{Utility function $u$}: We used (a) Entropic Risk Measure (ERM) with three different risk preference coefficients $\lambda=1.0$, $10.0$, $100.0$, and (b) Conditional Value-at-Risk (CVaR) with two different confidence levels $\alpha=0.9$, $0.95$.
\end{itemize}
We compared the following models:
\begin{itemize}
	\item \textbf{Proposed}: Adversarial deep hedging.
	\item \textbf{Brownian}: Deep hedging using the geometric Brownian motion as an underlying asset price process. The annual volatility parameter $\sigma$ is set to $0.2$.
	\item \textbf{Heston}: Deep hedging using the Heston model \cite{Heston1993} as an underlying asset price process. For the parameters of the Heston model, we followed the same setting as in \cite{Buehler2019b, Hirano2023-iiaiaai-ndh} (i.e., $\kappa = 1.0$, $\theta = 0.04$, $\rho = -0.7$, and $\sigma = 0.3$).
\end{itemize}

Additionally, we used the following parameters across all tasks: the exercise price of the option was set to 1.0, the option maturity was 20 days, the transaction cost ratio was $c=10^{-4}$, the learning rates for the hedger and the generator were $10^{-3}$, the two time-scale update rule \cite{Heusel2017} was set to G:H=1:5, the number of epochs was 5,000 for European option and 1,000 for Lookback option (due to computational resource constraints), 50,000 price series were used per epoch during training, and 10,000 price series were used per epoch during evaluation.
To stabilize the evaluation, we take the mean of 20 times trials only for the evaluation.

The adversarial learning interaction between the generator and the hedger can exhibit instability, which is a characteristic shared with GANs.
To address this, we adopted the performance indicator to take our model's best snapshot.
This is somewhat analogous to the use of the Inception score \cite{Salimans2016} or the Fréchet Inception Distance (FID) \cite{Heusel2017} in the GAN literature.
For our indicator, we employed the performance of the hedger of adversarial deep hedging when the underlying asset price process was Brownian.
However, we stress that it does not imply that the underlying asset price process was used during training.

We tested the best snapshot models of the hedger on real market data.
We used daily prices of the following three stock indices:
\footnote{Note that for simplicity, we used the Close-Close test, which differs from actual trading.}
\begin{itemize}
	\item S\&P 500: S\&P 500 (U.S.) data between January 2000 and August 2022
	\item S\&P 500 (old): S\&P 500 (U.S.) data between January 1931 and December 1950
	\item BVSP: Bovespa Index (Brazil) data between January 2000 and August 2022
\end{itemize}
We chose these data because they exhibit significantly different characteristics, such as market efficiency.
In general, emerging markets are considered to be more inefficient and exhibit characteristics such as long-term correlation \cite{Beben2001-correlations}.

We divided the evaluation data into time windows of $T = 20$ days each and tested the performance of the hedgers on each of these windows.
To align with the training conditions, we normalized the data such that the initial price $S_{t_0}$ of each time window was set to 1.0.
For each experimental setting, we reported the average and the standard deviation of performance across the windows.

\section{Results}\label{sec:result}
\begin{table*}[htbp]
	\centering
	\caption {Results of hedge prices (costs) for each setting. Underlined results are the best results among the proposed Brownian and Heston models. Bold results are better results than the Black--Scholes price.}
	\label{tab:result}
    \vspace{-2mm}
	S\&P 500\\
	\begin{tabular}{cc|c|cc|c}
		Option $Z$ & Utility $u$ & Proposed & Brownian & Heston & Black--Scholes\\\hline
		European & ERM(1.0) & $0.019669 \pm 0.000395$ & $\underline{\mathbf{0.018615 \pm 0.000228}}$ & $\mathbf{0.018720 \pm 0.000186}$ & $0.019143$\\
        European & ERM(10.0) & $0.020818 \pm 0.000162$ & $\mathbf{0.020619 \pm 0.000100}$ & $\underline{\mathbf{0.020578 \pm 0.000080}}$ & $0.020750$\\
        European & ERM(100.0) & $\mathbf{0.118480 \pm 0.002498}$ & $\underline{\mathbf{0.118334 \pm 0.001245}}$ & $\mathbf{0.120333 \pm 0.000959}$ & $0.123265$\\
		European & CVaR(0.9) & $\mathbf{0.055750 \pm 0.001051}$ &	$0.057322 \pm 0.000877$ & $\underline{\mathbf{0.052462 \pm 0.000888}}$ & $0.056678$\\
        European & CVaR(0.95) & $\mathbf{0.068432 \pm 0.000855}$ & $0.069850 \pm 0.000743$ & $\underline{\mathbf{0.063135 \pm 0.000953}}$ & $0.069778$\\
		Lookback & ERM(1.0) & $\mathbf{0.031467 \pm 0.001257}$ & $\mathbf{0.030218 \pm 0.000607}$ & $\underline{\mathbf{0.029981 \pm 0.000254}}$ & $0.031653$\\
        Lookback & ERM(10.0) & $\mathbf{0.035418 \pm 0.001309}$ & $\mathbf{0.034319 \pm 0.000264}$ & $\underline{\mathbf{0.034189 \pm 0.000250}}$ & $0.035609$\\
        Lookback & ERM(100.0) & $0.189564 \pm 0.009356$ & $\underline{\mathbf{0.169649 \pm 0.006907}}$ & $\mathbf{0.171893 \pm 0.003831}$ & $0.180721$\\
		Lookback & CVaR(0.9) & $\mathbf{0.084800 \pm 0.001601}$ & $\underline{\mathbf{0.082999 \pm 0.001293}}$ & $\mathbf{0.085991 \pm 0.002103}$ & $0.088665$\\
        Lookback & CVaR(0.95) & $\mathbf{0.107428 \pm 0.003131}$ & $\mathbf{0.106571 \pm 0.001689}$ & $\underline{\mathbf{0.102949 \pm 0.001497}}$ & $0.111228$\\
	\end{tabular}\\
	\vspace{2mm}
	S\&P 500 (old)\\
	\begin{tabular}{cc|c|cc|c}
		Option $Z$ & Utility $u$ & Proposed & Brownian & Heston & Black--Scholes\\\hline
		European & ERM(1.0) & $\underline{0.024419 \pm 0.000213}$ & $0.024681 \pm 0.000087$ & $0.024816 \pm 0.000065$ & $0.023999$\\
        European & ERM(10.0) &$0.027126 \pm 0.000151$ & $\underline{0.027073 \pm 0.000085}$ & $0.027075 \pm 0.000064$ & $0.027034$\\
        European & ERM(100.0) & $\underline{\mathbf{0.106246 \pm 0.002327}}$ & $\mathbf{0.109020 \pm 0.001406}$ & $\mathbf{0.110758 \pm 0.000523}$ & $0.111596$ \\
		European & CVaR(0.9) & $\mathbf{0.078522 \pm 0.001454}$ & $0.080840 \pm 0.001121$ & $\underline{\mathbf{0.074912 \pm 0.001011}}$ & $0.079936$\\
        European & CVaR(0.95) & $\mathbf{0.096169 \pm 0.001681}$ & $0.098889 \pm 0.001040$ & $\underline{\mathbf{0.090915 \pm 0.001360}}$ & $0.098017$\\
		Lookback & ERM(1.0) & $0.042866 \pm 0.000715$ & $\underline{0.042537 \pm 0.000191}$ & $0.042679 \pm 0.000170$ & $0.042158$\\
        Lookback & ERM(10.0) & $\underline{0.056832 \pm 0.003741}$ & $0.058438 \pm 0.001229$ & $0.059716 \pm 0.001132$ & $0.054546$\\
        Lookback & ERM(100.0) & $\underline{0.293896 \pm 0.010271}$ & $0.300853 \pm 0.026935$ & $0.300347 \pm 0.007946$ & $0.272272$ \\
		Lookback & CVaR(0.9) & $0.145088 \pm 0.002363$ & $0.142618 \pm 0.001593$ & $\underline{\mathbf{0.141590 \pm 0.004286}}$ & $0.141713$ \\
        Lookback & CVaR(0.95) & $\underline{0.179468 \pm 0.005448}$ & $0.188930 \pm 0.004209$ & $0.184107 \pm 0.002658$ & $0.177379$ \\
	\end{tabular}\\
	\vspace{2mm}
	BVSP\\
	\begin{tabular}{cc|c|cc|c}
		Option $Z$ & Utility $u$ & Proposed & Brownian & Heston & Black--Scholes\\\hline
		European & ERM(1.0) & $0.029468 \pm 0.000600$ & $\underline{0.028637 \pm 0.000352}$ & $0.028963 \pm 0.000280$ & $0.028476$\\
        European & ERM(10.0) & $0.031955 \pm 0.000250$ & $\mathbf{0.031765 \pm 0.000174}$ & $\underline{\mathbf{0.031721 \pm 0.000130}}$ & $0.031809$ \\
        European & ERM(100.0) & $\underline{\mathbf{0.191471 \pm 0.003796}}$ & $\mathbf{0.196748 \pm 0.001885}$ & $\mathbf{0.198085 \pm 0.000672}$ & $0.200292$\\
		European & CVaR(0.9) & $\mathbf{0.077644 \pm 0.002068}$ & $0.079088 \pm 0.001371$ & $\underline{\mathbf{0.071958 \pm 0.001465}}$ & $0.078619$\\
        European & CVaR(0.95) & $\mathbf{0.091748 \pm 0.001123}$ & $\mathbf{0.093617 \pm 0.001179}$ & $\underline{\mathbf{0.084048 \pm 0.001356}}$ & $0.093865$\\
		Lookback & ERM(1.0) & $0.049972 \pm 0.001463$ & $\underline{0.049670 \pm 0.000667}$ & $0.049796 \pm 0.000302$ & $0.048794$\\
        Lookback & ERM(10.0) & $0.056038 \pm 0.001958$ & $\underline{0.055078 \pm 0.000453}$ & $0.055470 \pm 0.000468$ & $0.054807$ \\
        Lookback & ERM(100.0) & $0.227208 \pm 0.013092$ & $\mathbf{0.195300 \pm 0.008136}$ & $\underline{\mathbf{0.194505 \pm 0.004359}}$ & $0.222651$ \\
		Lookback & CVaR(0.9) & $0.120657 \pm 0.002515$ & $\underline{\mathbf{0.117172 \pm 0.001533}}$ & $\mathbf{0.118355 \pm 0.003760}$ & $0.118696$\\
        Lookback & CVaR(0.95) & $\mathbf{0.138230 \pm 0.004309}$ & $0.142132 \pm 0.003011$ & $\underline{\mathbf{0.137341 \pm 0.001893}}$ & $0.139502$ \\
	\end{tabular}
\end{table*}

Table \ref{tab:result} shows the hedging test results on the actual data.
Each row corresponds to a different hedging criterion.
The values in the table represent the hedging cost $l(\delta)$, where lower values indicate superior hedging performance.
Underlined numbers correspond to the best results among three models compared, i.e., Proposed, Brownian, and Heston detailed in section \ref{sec:experiment}.
Bold numbers mean that the corresponding models outperform the Black--Scholes price.

According to the table, no single model is always better than others
For example, in the results on S\&P 500 data, the Heston model tends to outperform other models.
However, in the results on S\&P500 old data, our proposed model tends to outperform other models.

While our proposed model does not consistently outperform the other baseline models, the differences between our model and others are relatively minor, typically less than 2\%.

\begin{figure*}[htb]
    \begin{minipage}[b]{0.33\linewidth}
    	\centering
    	\includegraphics[width=\linewidth]{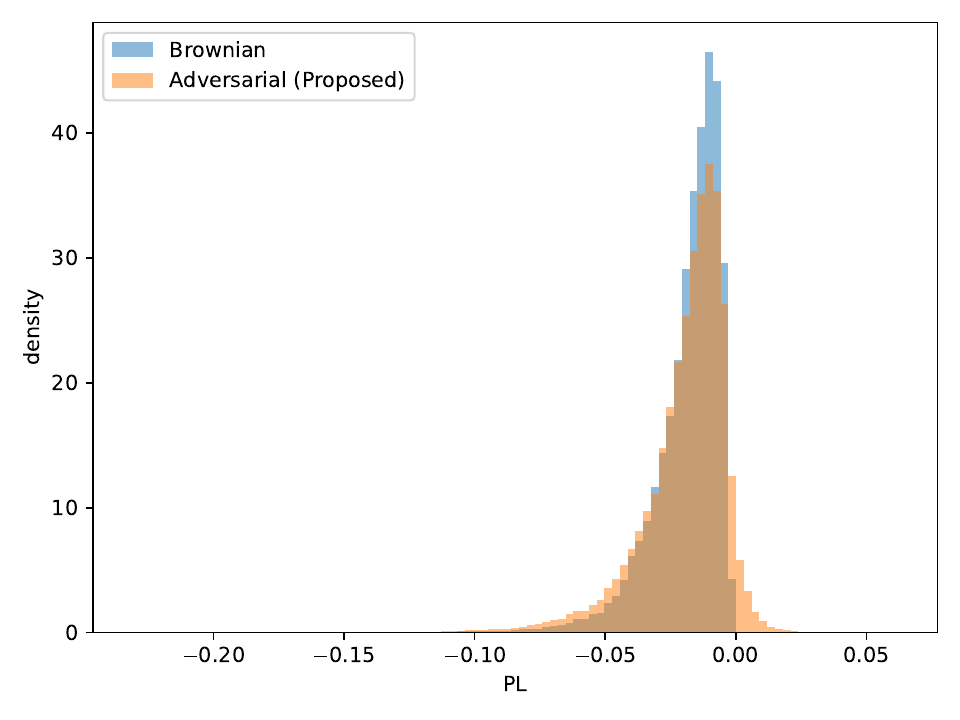}
    	\subcaption{ERM(1.0) \& European Option}
    	\label{fig:european-erm}
    \end{minipage}
    \begin{minipage}[b]{0.33\linewidth}
        \centering
        \includegraphics[width=\linewidth]{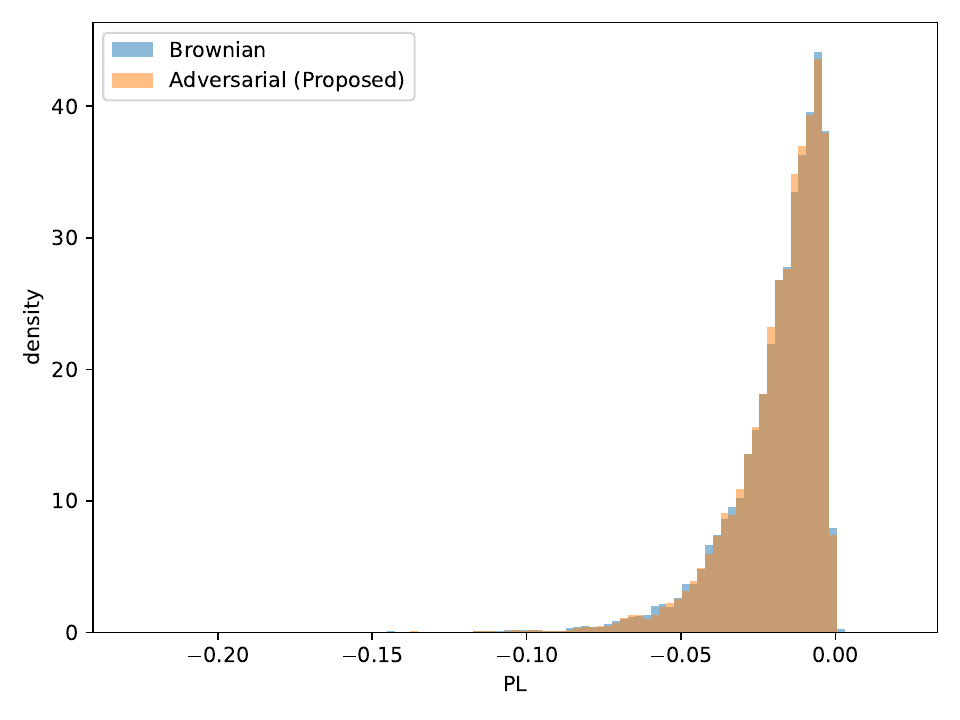}
        \subcaption{CVaR(0.90) \& European Option}
        \label{fig:european-cvar}
    \end{minipage}\\
    \begin{minipage}[b]{0.33\linewidth}
        \centering
        \includegraphics[width=\linewidth]{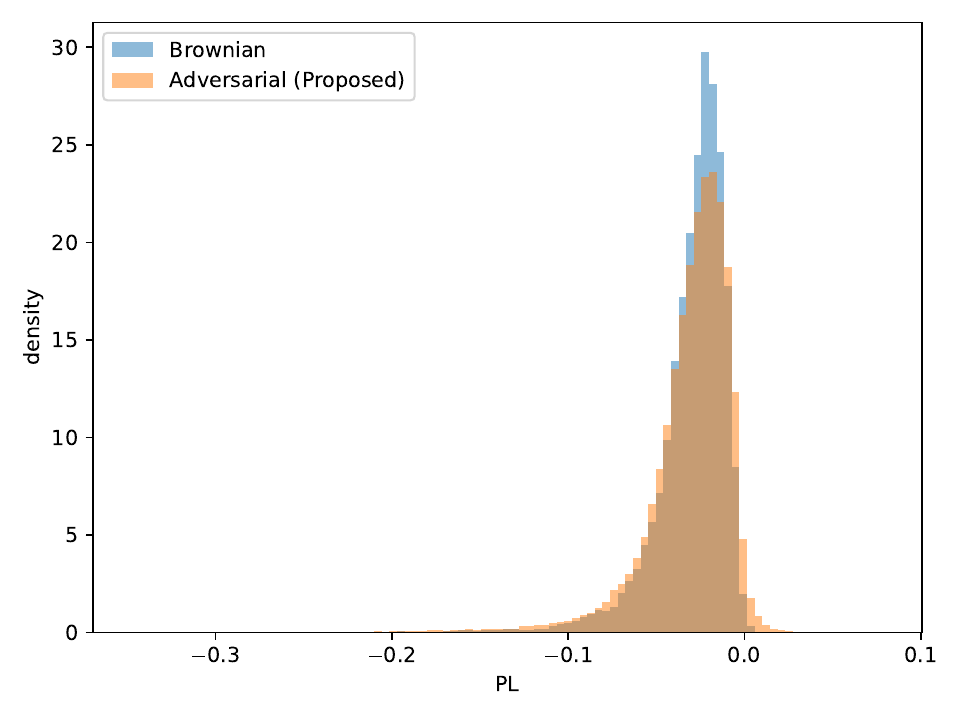}
        \subcaption{ERM(1.0) \& Lookback Option}
        \label{fig:lookback-erm}
    \end{minipage}
    \begin{minipage}[b]{0.33\linewidth}
    	\centering
    	\includegraphics[width=\linewidth]{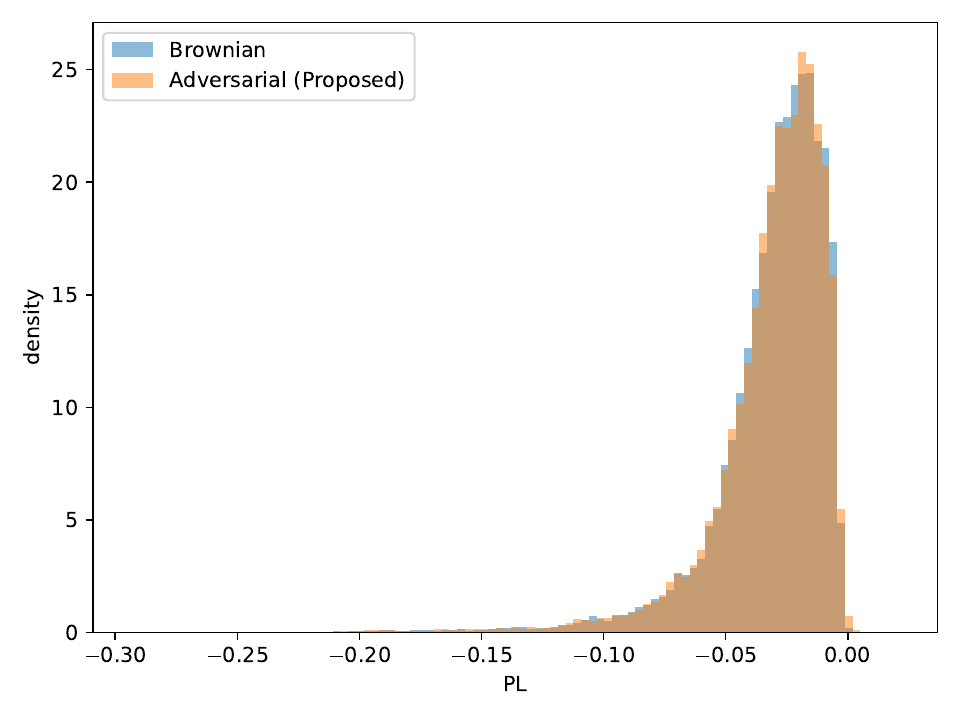}
        \subcaption{CVaR(0.90) \& Lookback Option}
    	\label{fig:lookback-cvar}
    \end{minipage}
    \caption{Histograms of PL for Brownian (blue) and the proposed (orange) method. The utility function and option type differ in each figure. Backtesting data is S\&P500.}
\end{figure*}

Figure \ref{fig:european-erm} -- \ref{fig:lookback-cvar} show the distribution of the PL of hedging results on the S\&P 500 data backtest.
The Brownian model and our proposed model are represented by the blue and orange histograms, respectively.
According to these results, we can observe almost all of the PL distributions achieved by the Brownian model are replicated by our proposed adversarial deep hedging.
This replication is especially noticeable when the CVaR is used as the utility function (Figure \ref{fig:european-cvar} and Figure \ref{fig:lookback-cvar}).
Interestingly, it is apparent that when the utility function is ERM, the proposed model generates positive PL with a positive probability.
This suggests that the adversarial learning might be encouraging the hedger to explore a more diverse range of opportunities to attain profits.

\section{Discussion}\label{sec:discussion}
\subsection{Performance of Adversarial Deep Hedging}
According to the experimental results, the proposed approach does not consistently outperform the baselines. However, it does achieve promising performance.
Regarding the S\&P500 data, the Heston model tends to outperform other models. 
In contrast, for the S\&P500 old data, the proposed approach shows a higher likelihood of outperforming other models.
Nonetheless, what we want to emphasize is that the proposed model achieves comparable performance to other models, regardless of the evaluation data or the utility function.
This result is noteworthy because the proposed method does not rely on explicit modeling of underlying asset price process.

In general, the optimal choice of the underlying asset model is not trivial and depends on the utility function and the market data used for testing.
In the experimental results, when the utility function is ERM, the Brownian model tends to perform relatively better than the Heston model.
On the other hand, when the utility function is CVaR, which focuses more on tail risk, the Heston model tends to achieve higher performance.
Moreover, there are calibration issues with the underlying asset models. Traditionally, parameters for mathematical finance models are estimated to fit the historical market data or the implied volatility.
We should also note that, with more complex models such as rough volatility models, the calibration itself can be unstable or computationally challenging (see e.g., \cite{Horvath2021-deep-learning-volatility}).
However, the training of deep hedging requires the generation of future scenarios through simulations, and the standard calibration methods may not necessarily align with this purpose.
As the Brownian model and the Heston model have only a few parameters, they might not be well-suited for simulating future paths in highly non-stationary real-world markets. This is also suggested by the high performance of the proposed method in S\&P500old and BVSP, which are considered as more non-stationary markets.
One possible reason for this is that the adversarial learning can train the hedger efficiently while avoiding overfitting to historical data.
Considering these points, the proposed method has the potential to be more generally useful across broader markets and utility functions.

\subsection{Generator Volatility}\label{sec:discussion-vol}

\begin{figure}[tbp]
	\centering
	\includegraphics[width=0.8\linewidth]{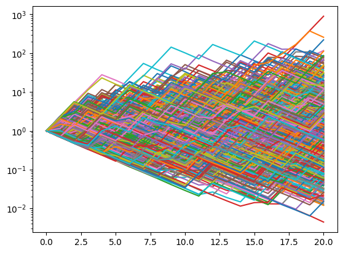}
	\caption{Generated price path example.}
	\label{fig:generated_path}
\end{figure}

Figure \ref{fig:generated_path} shows an example of paths generated by the generator learned through adversarial learning.
The distribution of paths appears roughly symmetric in the logarithmic domain and qualitatively resembles a discretized geometric Brownian motion.
However, the generated paths exhibit significantly large volatilities.
This can be attributed to the fact that, as discussed in Section \ref{sec:num-analysis}, the utility function can monotonically decrease with an increase in the generator's volatility, leading to the exceedingly high volatility of the adversarially learned generator.
Nevertheless, in our experiments, the hedger seems to have been effectively trained.
One possible explanation is that hedging European options and lookback options possess a certain scale invariance, where volatility has a relatively minor impact on the optimal hedging actions.
In particular, when the strike $K$ and initial value $S_0$ are the same, the payoff of a European option $\max \{ S_T - S_0, 0 \}$ is \textit{positive homogeneous}.
Indeed, if the generator volatility is multiplied by a constant factor $\alpha$ (thus the random variable $S_T - S_0$ as well), the probability of the payoff occurring remains unchanged, and the value of the payoff simply becomes $\alpha$ times the original value.
As a result, the optimal hedging action remains invariant under such scaling of volatility.
However, when dealing with financial derivatives whose optimal hedging actions depend on the path's scale, such as volatility swaps or barrier options, it may be necessary to impose certain volatility constraints during training.
Developing effective volatility constraints (or regularizations) is left for future work.
Additionally, in practical derivative businesses, underlying asset models are commonly calibrated to observed volatility surfaces in the market.
Developing techniques for adversarial deep hedging to calibrate the generator to real volatility surfaces would also be an interesting future work.

\section{Conclusion}\label{sec:conclusion}
In this study, we proposed a novel approach named adversarial deep hedging, aimed at efficient training of deep hedging strategies without reliance on an assumed underlying price process.
The key idea is to incorporate the concept of adversarial learning.
We introduce a new component, the generator, that is responsible for generating the price process, and the hedger and the generator engage in adversarial learning through a min-max game.
The key advantage of employing adversarial deep hedging lies in its ability to obviate the need for explicit modeling of underlying price processes, thereby circumventing any learning biases that may arise from modeling assumptions.
In the experiments, we empirically demonstrated that the proposed method exhibits competitive hedging performance even in the absence of an explicit price process model.

There are a number of future research directions suggested by this study.
The adversarial deep hedging tends to diverge the generator volatility. Therefore, it is essential to explore appropriate constraints or regularization on volatility without compromising the hedge performance. Additionally, as adversarial deep hedging currently incurs significant computational costs, investigating more efficient learning algorithms is also a crucial task.

\bibliographystyle{ACM-Reference-Format}
\bibliography{cite}

\end{document}